\documentclass[pdftex,twocolumn,10pt,letterpaper]{article}
\usepackage{graphicx, times}
\usepackage{lipsum}
\usepackage{listings}
\usepackage{xcolor}

\definecolor{codegreen}{rgb}{0,0.6,0}
\definecolor{codegray}{rgb}{0.5,0.5,0.5}
\definecolor{codepurple}{rgb}{0.58,0,0.82}
\definecolor{backcolour}{rgb}{0.95,0.95,0.92}

\lstdefinestyle{mystyle}{
    backgroundcolor=\color{backcolour},   
    commentstyle=\color{codegreen},
    keywordstyle=\color{magenta},
    numberstyle=\tiny\color{codegray},
    stringstyle=\color{codepurple},
    basicstyle=\ttfamily\footnotesize,
    breakatwhitespace=false,         
    breaklines=true,                 
    captionpos=b,                    
    keepspaces=true,                 
    numbers=left,                    
    numbersep=5pt,                  
    showspaces=false,                
    showstringspaces=false,
    showtabs=false,                  
    tabsize=2
}

\begin{document}
\title{Create Benchmarks for Data Lakes}
\author{Yi Lyu, Pei-Chieh Lo, and Natan Lidukhover}
\date{}

\interfootnotelinepenalty=10000

\maketitle

\section{Introduction}

% 1-2 paras: what is the problem you are solving, and why is it important (need citations)

The problem that we are trying to solve is to build a new benchmark for the data lakes. Although there is an increasing need for data lakes these days, there are not many standardized benchmarks for data lakes. By doing that, we are hoping to get a more objective and comparative understanding of different data lake implementations.

There are many users requiring a database system that offers more flexibility on storage and analytics than traditional data warehouses\cite{AWS} \cite{Snowflake}\cite{efflex}\cite{vetrass}\cite{catp}\cite{monom}\cite{zhang2021tapping,zhang2023first,feng2021allign,feng2024f3,feng2025optimus,han2022francis,ctf,safeguard}. A data lake is a large repository for storing and analyzing data without the restriction on the data structure, type, and size. It is an increasingly popular way to store  data because it allows users to manipulate different data types from different sources in a centralized repository. Therefore, having standard benchmarks for data lake systems can assist in the process of developing better data lakes that take performance metrics into account. 

The data lake market was valued at \$3.74 billion in 2020\cite{Mordor}. By 2026, this market is expected to reach \$17.6 billion, meaning the compound annual growth rate of this market over the period from 2021 to 2026 is 29.9\%. This is a rapidly growing market, and according to the O'Reilly Data Scientist Salary Survey, the speed of data retrieval is one reason why data lakes are preferred and growing more when compared to a traditional data warehouse. Given that speed is vital for data lakes, it is increasingly important to have a benchmarking suite that is capable of determining the performance of a given data lake implementation to see which ones are faster and more performant.

The top data lakes in the market are owned by the biggest cloud computing platforms: Amazon Web Services, Microsoft Azure, Google Cloud Platform, and Oracle Cloud Infrastructure\cite{TopDataLakes}. It will be useful to compare at least some of these to one another in terms of performance and will provide some immediate use for our benchmark. Furthermore, as mentioned in Related Work, DLBench is currently a data lake benchmark that claims to be the first one, and it ran its testing on an open-source data lake called AUDAL. That would be another useful data lake to compare to as a baseline for how our suite compares to DLBench.

% 1-2 paras: How other people solve and why they fall short
Currently, there has not been a lot of work on data lake benchmarks. Existing benchmarks for other big data systems cannot be directly applied here. The traditional benchmarks such as the TPC standard specification \cite{TPC-H} only provide SQL workloads and only work for structured data stores. However, data lakes support structured data, unstructured data, and semi-structured data. Also, there is one operation that is seldom tested in existing benchmarks: finding similarity. This is a common operation in data lakes, so a benchmark should also do performance testing for this operation.

Because data lakes can accept many types of structures for data, it is important to understand the differences between structured, semi-structured, and unstructured data. Structured data has a pre-defined structure and is formatted that way before being placed in storage. An example of this is tabular data inputted into a relational database such as SQL. SQL data follows a set schema and is therefore structured. Unstructured data, on the other hand, is not processed until it is taken out of the data store. An example of this is raw text data or video files. Semi-structured data is a bit more confusing to understand. It is a middle state between raw data and strictly typed fields such as those in a SQL schema. An HTML file is an example of this type of data, as tags and anchors provide a rigid structure that can be fed into a parser and have the explicit structure extracted; however, many HTML pages contain text and images, and other raw data that cannot be parsed out. For these reasons, it is not entirely processed on input into a data store nor on reading out, but rather both times, and it is semi-structured data\cite{SemiStructured}.

% \cite{DBLP:journals/corr/abs-2110-01227}

% 1-2 paras: How do you plan on solving it and why your approach is better

We plan to have more comprehensive benchmarks for data lakes by (1) extending the target data structure type to structured, semi-structured, and unstructured, (2) measuring time for different workload model examples  including data retrieval, data aggregation, and data queries, and (3) observing the performance with a variety of data access operations including query execution time, metadata generation time, and metadata size. The approach we are planning to take is more inclusive and will take more aspects into account in order to figure out what will be the most critical part for data lake systems in storing data and accessing it. 

% 1 para:Anticipated results or what experiments you will use

The expected results will contain scripts to generate data with different scale factors which others can run on their own to get the performance metrics from a given data lake implementation. We would also run this script on our own to get the measuring number for each metric under different types of structured data. We would like to get some insights from the statistics and highlight our new findings for data lakes benchmarks. We will use CloudLab for our experiments.
 
\section{Related Work}
% \lipsum[2-5]
Currently, there are still no standardized benchmarks to benchmark the performance of data lakes. TPC-H\cite{TPC-H} and TPC-DS\cite{TPC-DS} issued by the Transaction Processing Performance Council are still widely adopted by the big data system community. 

%These benchmarks creates workloads for traditional databases, that means they only have SQL workloads. Also they only do testing for structured data.

BigBench\cite{BigBench} is also another popular benchmark that executes SQL queries to evaluate the performance of data systems. In addition to only structured data, it also includes more complex big data analysis tasks—namely sentiment analysis over short texts.

All the benchmarks mentioned above are not a great fit to benchmark data lakes. For TPC-H\cite{TPC-H} and TPC-DS\cite{TPC-DS}, they are only for traditional databases. That means they only have SQL workloads. Also, they only do testing for structured data. But data lakes store structured data, unstructured data, and semi-structured data at the same time. These are not enough to test all functionalities of data lakes.

AdBench is a benchmark for modern end-to-end data pipelines that integrate with data lakes \cite{AdBench}. AdBench notes that with the emergence and increasing utilization of Apache Hadoop Distributed File System for data lake implementation and construction, data pipelines are being built on top of this platform as well. AdBench has expanded on previous standalone task benchmarks such as ones for Analytical SQL queries. AdBench specifically runs on ad-serving and streaming data, but the paper notes the characteristics of these workloads are analogous to those in Internet of Things (IoT) infrastructure, healthcare, and financial services, meaning the concept can be extended to those industries if needed. AdBench does not directly serve as a data lake benchmark, but it provides a useful set of metrics that compares across scale factors, just as we wish to do, in a closely related area to data lakes: the end-to-end pipelines built on top of data lakes.

For BigBench\cite{BigBench}, it does a better job of testing not just structured data, but also unstructured data (text files). But it still lacks support for semi-structured data.

Similar to BigBench, there is a relatively new proposed data lake benchmarking system known as DLBench\cite{DLBench} that runs tests using structured and unstructured data. DLBench is limited to running on data lake implementations that support text and table content, specifically referring to a data model comprised of text and CSV documents. DLBench ran testing and captured performance metrics specifically on an open-source data lake system previously developed by the authors, so its results are skewed by that fact. DLBench should be tested on more widely used data lake systems to better display its efficacy.

According to the paper, DLBench claims to be the first data lake benchmark. There is some credence to this statement given the fact that DLBench is one of the few benchmarking suites that is able to run on structured and unstructured data. Furthermore, DLBench ran test suites on both data models and query models, which is something we also wish to perform. The examples given in DLBench were run on the AUDAL data lake implementation. The paper found that metadata size and generation time both scaled exponentially on AUDAL as the scaling factor increased. This is one of many useful results in that paper we will try to emulate.

\section{Approach}
Our approach to solving the problem is to design a standardized data lake benchmark to compare different data lake frameworks together.

For the source dataset, we are considering using structured data, unstructured data, and semi-structured data to simulate all possible scenarios in data lakes.

For structured data, we choose table structured data in CSV format since it is the standardized data storage format in relational databases, and we used the IMDb movie dataset for this. For semi-structured data, we are using XML data. XML is a great protocol for storing, transmitting, and reconstructing data, and it gains high popularity for its high readability. For unstructured data, we chose a simple text file of Apache HTTP logs.

\begin{figure}[h]
    \centering
    \includegraphics[width=6cm]{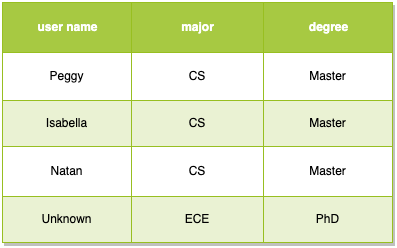}
    \caption{Example of structured dataset}
    \label{fig:workflow}
\end{figure}

\begin{figure}[h]
    \centering
    \includegraphics[width=6cm]{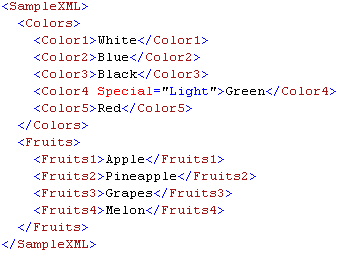}
    \caption{Example of semi-structured dataset}
    \label{fig:workflow}
\end{figure}

\begin{figure}[h]
    \centering
    \includegraphics[width=6cm]{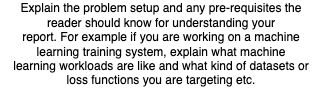}
    \caption{Example of unstructured dataset}
    \label{fig:workflow}
\end{figure}

For the query workload generation, we are using both single-keyword and multi-keyword searches.

After a discussion with Song, we agreed on the fact that our time is too limited to deploy it on real data lake frameworks. Thus, we choose to implement the EASE\cite{EASE} algorithm on a local machine as the test bed. The EASE keyword search algorithm can be applied to all different data formats by first modeling them into graphs. Then, EASE constructs graph indices instead of traditional inverted indices.

\section{Design}
The system architecture of our benchmark framework is listed below.

\begin{figure}[h]
    \centering
    \includegraphics[width=4cm]{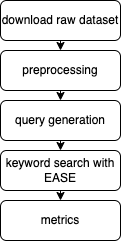}
    \caption{Overview of the benchmark system}
    \label{fig:workflow}
\end{figure}

The benchmark system starts by downloading the raw dataset. After much effort, we collected data from IMDb (Internet Movie Database)—a movie rating CSV dataset (structured data), DBLP (Database systems and Logic Programming)—a bibliography XML dataset (semi-structured data), and Apache HTTP logs—a text dataset (unstructured data).

% Need some introduction about each dataset and a conceptual graph

DBLP is a computer science bibliography website that listed more than 5.4 million journal articles, conference papers, and other publications on computer science\cite{DBLP}. We downloaded the public dataset of DBLP (2019-04-01) in XML format. 

\begin{figure}[h]
    \centering
    \includegraphics[width=9cm]{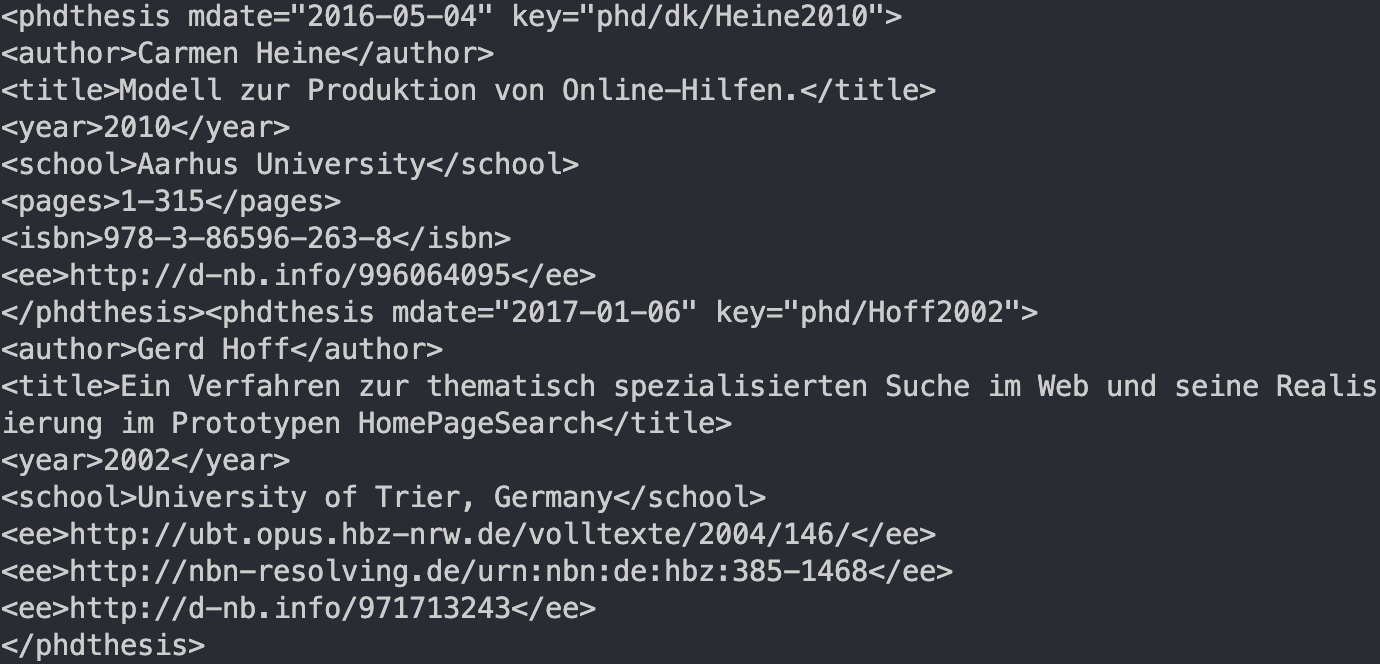}
    \caption{Example of raw DBLP XML dataset}
    \label{fig:DBLP_dataset}
\end{figure}

It includes all the papers that were listed in DBLP at that time and each individual record shows the author, title, and other information, as shown in figure \ref{fig:DBLP_dataset}. 

Followed by that, we preprocessed all the datasets in order to turn them into the nodes and edges required by EASE. For EASE, we need to input both a node file and an edge file. For nodes, we provide the keywords of each separated by space. For edges, we just simply give two node indices that indicate they are related.

% How to do preprocessing for each individual data types

For the structured dataset, we chose the IMDb dataset that contains multiple CSV files. The files we selected as the preprocessing input are as below:

\begin{itemize}
    \item movie.csv: It contains the basic information for each movie including the genres, year, and movieId.
    \item rating.csv: It includes the information about which userId rates which movieId and the rating itself
    \item tags.csv: It indicates the tags that the userId gave for movieId
\end{itemize}

We defined there is a relationship between movieId and userId if the user ever rated or tagged the movie. And the nodes for this dataset are the userId and movieId that ever exist in any file. And the keywords for the movie node would include the genres, the year, and the tags of the movie. For the user node, its keywords would be the movie title that it was related to. We would parse each file and get the keywords and relationship for each node and then output as the node and edge file for EASE.

\begin{figure}[h]
    \centering
    \includegraphics[width=8cm]{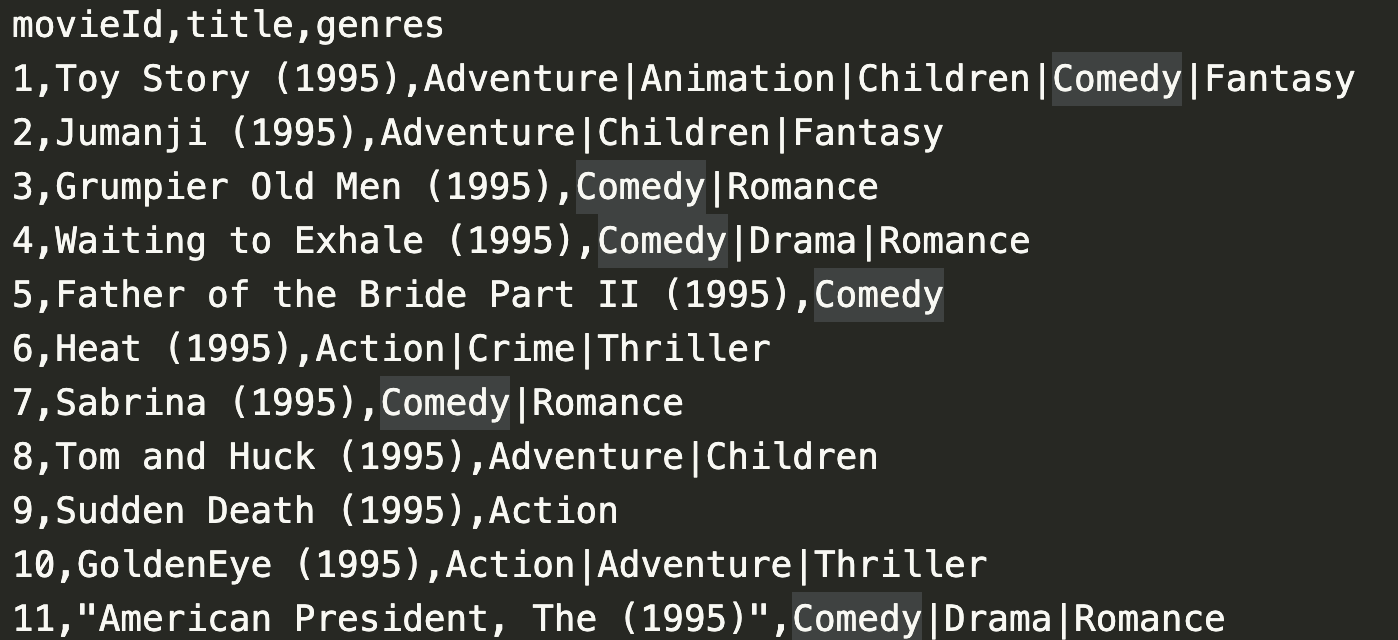}
    \caption{Example of raw movie structured dataset}
    \label{fig:DBLP_dataset}
\end{figure}

The unstructured dataset was raw text files of Apache HTTP logs taken from an open-source dataset \cite{Apache}. The preprocessing step was a bit tricky as we had to define what part of this text data would be the node and what would represent the edges. The nodes were made up of the actual log text messages that were present in these files. This meant the message after the date and IP address that were present on each line. The edges were a mapping of the time (hour and minute concatenated) that each log message was printed to the nodes (represented by a logLineId).

The following is an example of a few lines from the text dataset:

\begin{lstlisting}[language=bash]
"192.168.4.25 - - [22/Dec/2016:16:30:52 +0300] "POST /administrator/index.php HTTP/1.1" 303 382 "http://192.168.4.161/DVWA" "Mozilla/5.0 (Windows NT 6.1; WOW64) AppleWebKit/537.21 (KHTML, like Gecko) Chrome/41.0.2228.0 Safari/537.21""
"192.168.4.25 - - [22/Dec/2016:16:29:05 +0300] "POST /index.php/component/search/ HTTP/1.1" 500 2011 "-" "Mozilla/5.0 (Windows NT 6.1; WOW64) AppleWebKit/537.21 (KHTML, like Gecko) Chrome/41.0.2228.0 Safari/537.21""
\end{lstlisting}

For the XML dataset, we let each paper and its authors be separate nodes. If a person is one of the paper's authors, we build an edge.

After that, we generated keyword queries for the workloads. After trying multiple solutions, we finally decided to generate the keyword search with word statistics.

\begin{figure}[h]
    \centering
    \includegraphics[width=7cm]{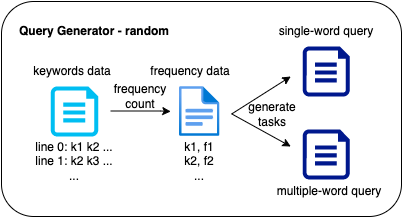}
    \caption{Query generator architecture: Random case}
    \label{fig:my_label}
\end{figure}

\begin{figure}[h]
    \centering
    \includegraphics[width=7cm]{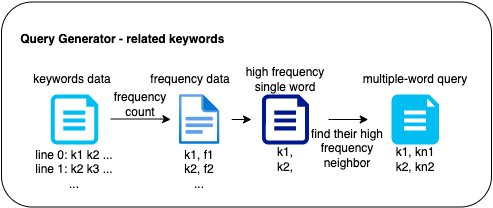}
    \caption{Query generator architecture: Related case}
    \label{fig:my_label}
\end{figure}

First, we counted each different word's frequency in the node.csv file. By doing that, we get all the keyword rankings from the highest frequency to the lowest frequency. The intuition for doing that is that we need the existence of an r-radius graph to execute keyword search queries using the EASE system. The r-radius graph is the subgraph each node has with its at most r hops neighbors. So we want more edges when possible.

According to the power law, most nodes do not have a large number of neighbors. In order to remove the keywords that may not have many edges, we are going to abandon the bottom 95\% keywords and narrow our search space to the top 5\%.

The user can then specify how many single-word queries and how many multiple-word queries they want to have. 

We are going to generate two different sets of queries: single-keyword search queries and multiple-keyword search queries.

For single-keyword search queries, our automatic script will randomly pick the number of words from the top 5\% keywords in our search space and finally generate the query file.

For multiple-keyword search queries, we want to first test the performance of random connection nodes and also highly connected nodes. For random connection nodes, we generate queries similar to single-keyword search queries. We repeat the process multiple times to pick out a certain number of keywords from the top 5\% of keywords.

In order to find highly connected nodes (nodes with multiple neighbors), we first find the first keyword purely with its frequency. In order to keep the node highly connected, we find the neighbors of the chosen keyword within a radius, r. Within that, we are going to find the keyword with the highest frequency. If there are more neighbors, we keep doing that, and if not, we will move to a new keyword.

\begin{figure}[h]
    \centering
    \includegraphics[width=6cm]{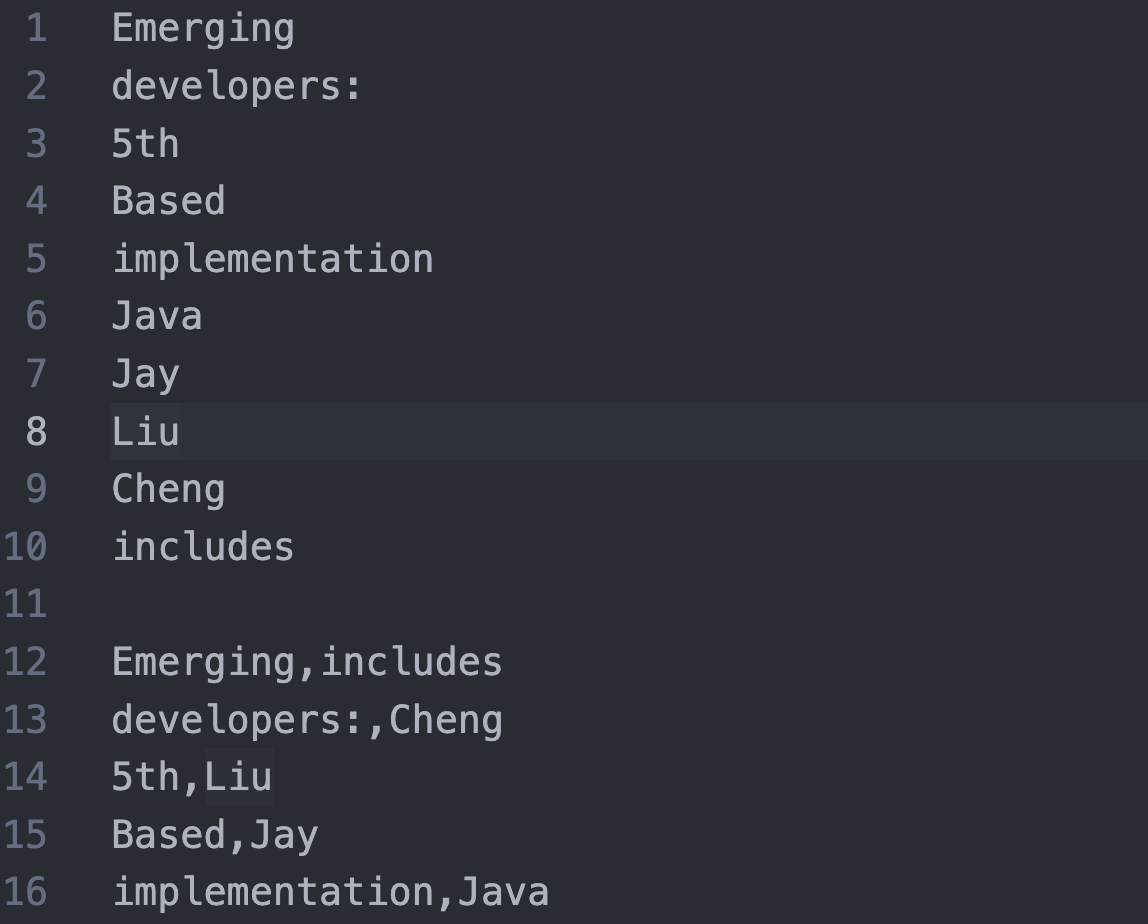}
    \caption{An overview of the query file}
    \label{fig:query}
\end{figure}

As shown in figure \ref{fig:query}, the query file has the keywords of each query on different lines. For lines 1-10, there are 10 different queries. We combine the single-keyword search queries in a test suite. Lines 12-16 are examples of multiple-keyword searches. They also constitute a test suite.

\section{Evaluation}
\subsection{Experiment Setup}
For evaluation, we are using a MacBook Pro with 16GB memory and 512 GB disk. The chip is an M1 Pro and the OS is macOS Monterey Version 12.1.

\begin{table}[]
    \centering
    \begin{tabular}{|c|c|}
        \hline
        Brand & Apple \\ \hline
        OS & macOS Monterey (12.1) \\ \hline
        Memory & 16GB\\ \hline
        Disk & 512 GB \\ \hline
        Chip & M1 Pro \\ \hline
    \end{tabular}
    \caption{Evaluation environment configuration}
    \label{tab:my_label}
\end{table}

In order to evaluate our system, we are going to compare the graph building time, single-keyword query execution time, and multiple-keyword query execution time with different numbers of papers. 

The dataset we chose was XML data. Here, we chose paper numbers 2000, 4000, 8000, 16000, 32000, and 64000. We wanted to see how the execution time for the dataset would change when the scale of the dataset was doubled.

For single-keyword query generation, we are going to pick 10 random keywords from the top 5\% most frequent keywords. For multiple-keyword query generations, we are going to generate two different queries, one with 5 keywords and one with ten keywords.

\subsection{Experiment Result}
In this section, we are going to introduce the results of our experiments and also list our findings after doing the experiments.

\subsubsection{Graph Building Time}
\begin{figure}[h]
    \centering
    \includegraphics[width=8cm]{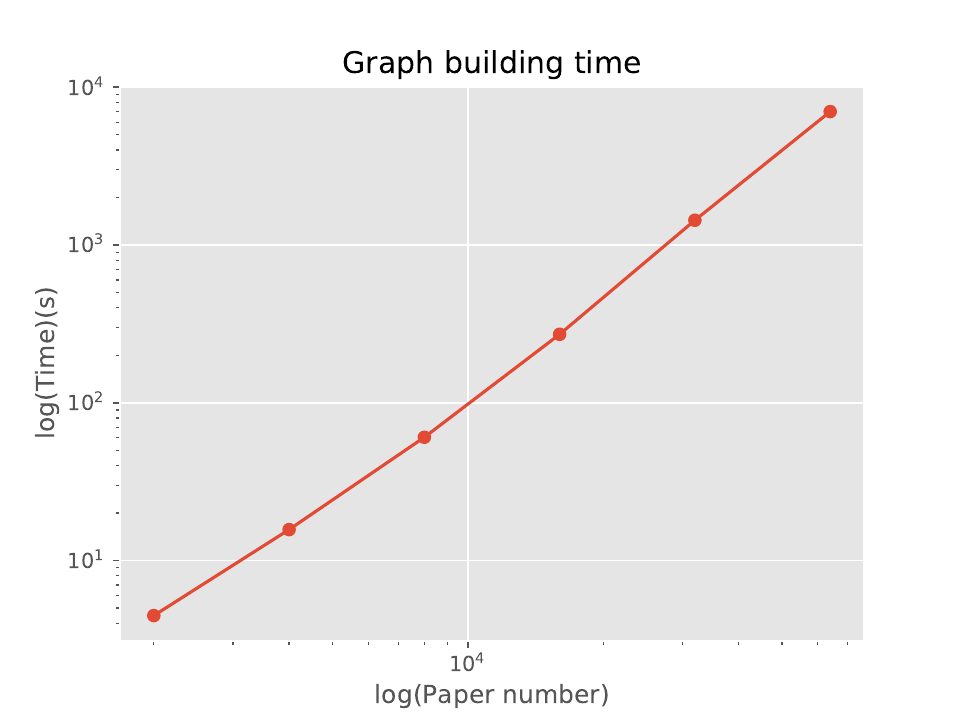}
    \caption{Graph building time with dataset doubling}
    \label{fig:graphBuildingTime}
\end{figure}

According to the design of our testing algorithm EASE\cite{EASE}, it first creates a graph with all the nodes and then executes the queries on it. So, we start our experiments by exploring the relationship between dataset size and graph building time.

As we can see from figure \ref{fig:graphBuildingTime}, when the dataset size doubles, the graph building time increases almost linearly. That means that the distribution of the dataset is even and sparse. With the addition of new nodes, it does not create many new edges with existing nodes.

\subsubsection{Single-Keyword Query Execution Time}
After testing the graph building time, we are going to explore how the single-keyword search query test suite(contains 10 different keyword queries search). The execution time listed below only records the time from starting the keyword search query to getting keyword search results(excluding graph building time).

\begin{figure}[h]
    \centering
    \includegraphics[width=8cm]{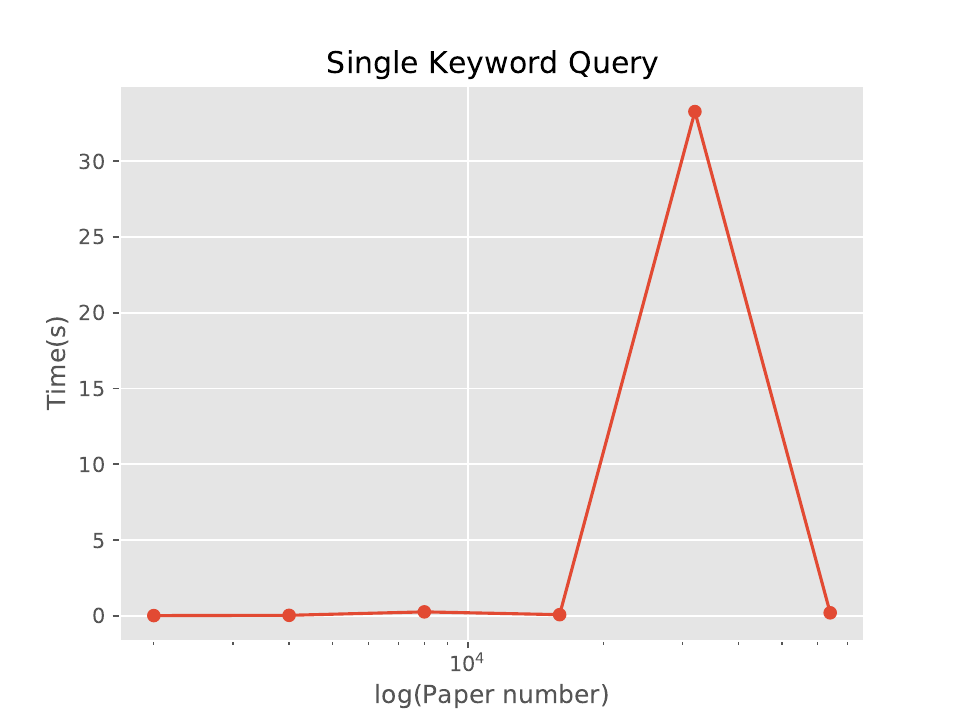}
    \caption{Single-keyword queries execution time with dataset doubling}
    \label{fig:singleKeywordQueryTime}
\end{figure}
According to the result above\ref{fig:singleKeywordQueryTime}, we notice that there is an outlier in the graph. When the dataset size hits 32000, the query time increases greatly (around 33s), while the query times for other dataset sizes remain less in a second. The reason for the outlier could be the structure of the graph. When the keyword that we are searching for has too many neighbors and subgraphs, it will take more time to execute the queries. In order to better analyze the more general trends when the dataset size doubles, we remove the outlier and get a new graph below.

\begin{figure}[h]
    \centering
    \includegraphics[width=8cm]{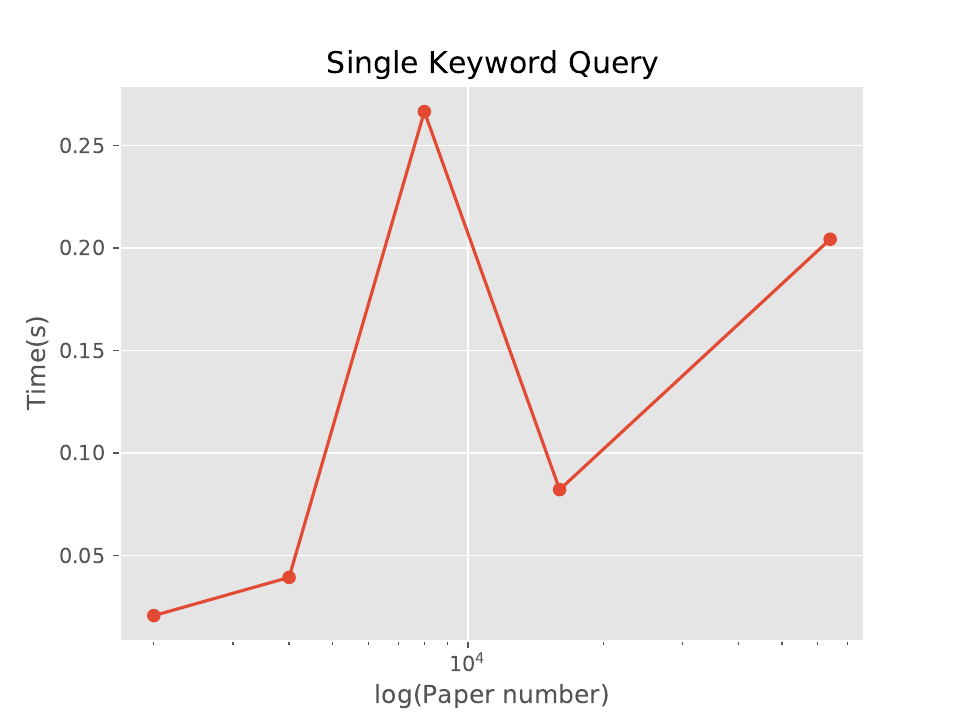}
    \caption{Single-keyword queries execution time with dataset doubling (Except straggler)}
    \label{fig:singleKeywordQueryTimeWithoutStraggler}
\end{figure}

Although there are some fluctuations in the graph, the query execution time increases when the size of the dataset increases. Starting around 0.02s, it reaches around 0.2s when the dataset size increases by 32 times. One reason for that is that the graph is greatly expanded when the dataset size increases. There is a larger search space and we have more subgraphs to examine. However, the execution time does not grow by the same amount as the dataset. It is because the EASE algorithm splits the graphs into subgraphs and we can do pruning without visiting all nodes. That makes the overall time complexity lower than the size of the entire dataset.

\subsubsection{Multiple-Keyword Query Execution Time}
As mentioned before, we also removed the outlier when the dataset size hit 32,000.

\begin{figure}[h]
    \centering
    \includegraphics[width=8cm]{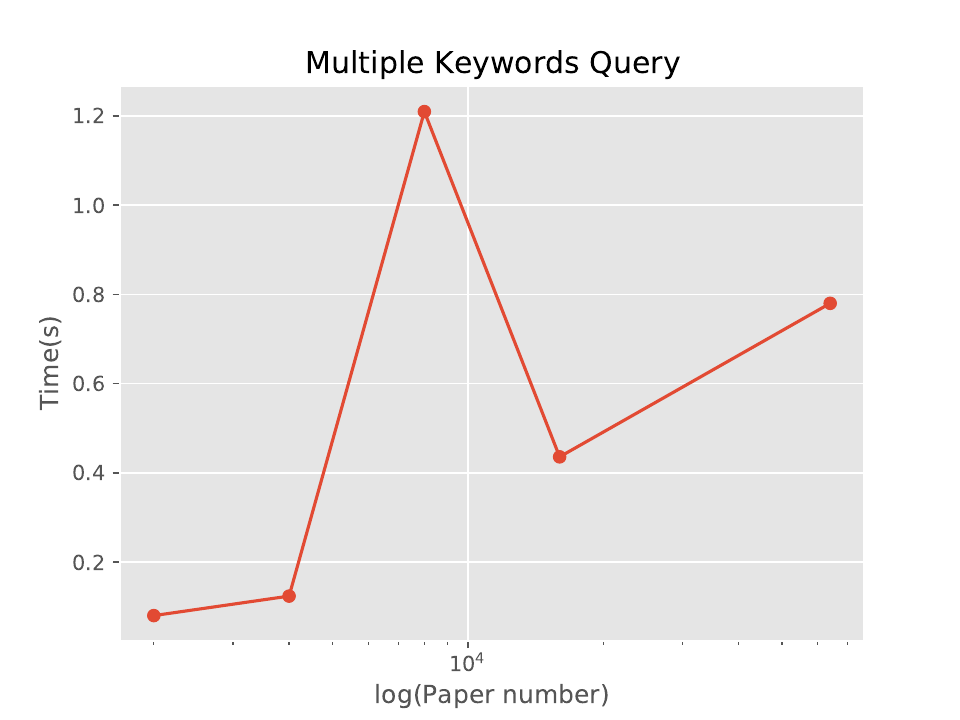}
    \caption{Multiple-keyword queries execution time with dataset doubling (Except straggler)}
    \label{fig:multipleKeywordQueryTimeWithoutStraggler}
\end{figure}

From the above graph, we find similar trends with single-keyword queries. But the time of multiple-keyword query execution is greatly higher than single-keyword query execution. The reason for that is that we have more requirements for the subgraph and there are more things to examine. With more keywords, the system will spend more time.

\section{Conclusion \& Future works}

In this paper, we presented a new benchmark for evaluating data lakes that can store all types of datasets including structured, semi-structured, and unstructured datasets. We described the methods that we used to create the benchmark, including data collection, data preprocessing, query generation, and the keyword search algorithm: EASE. 

For the data collection part, we collected the data from the IMDb, DBLP, and Apache HTTP logs datasets. Then, we preprocessed the data with corresponding scripts that can then get the nodes(keywords) and edges(relationship) files.

For the query generation part, we established Python scripts to do the word frequency count and output the single-word query and multiple-word query with either random or related policies. We modified an existing keyword search algorithm implementation, EASE, on our testbed. 

We conducted several experiments with our benchmark to evaluate the data lake model we implemented in order to see the performance. 

From the experiment results, we found the below findings with our test data lake:

\begin{itemize}
    \item Although there are some fluctuations, the query execution time increases when the size of the dataset increases.
    \item The time of multiple-keyword query execution is greatly higher than single-keyword query execution.
    \item The graph building time, which is also the data lake formation time, increases almost linearly when the dataset size doubles.
\end{itemize}

Although we completed the test and experiment for the test data lake, there are still some limitations in our approach. 

The query generation algorithm can be more diverse and personalized, and output more queries. It would be great to add the feature that users can have options for their own queries. That way, users could test the tasks that they focused on. 

Also, we would like to see if our benchmark can test on real data lake frameworks. Due to the limited time we have in this class, we did not have time to set up a real data lake framework, as discussed with Song. If we had a chance to do more in the future, it would be great to see our benchmark run with real-world test cases using actual data lake implementations.

Overall, our benchmark represents a possible solution for data lakes that can run keyword search tasks to test their performance, and we hope that it will serve as a useful tool for evaluating and improving the performance of data lakes.

{
\bibliographystyle{plain}
\bibliography{ref}
}
\end{document}